\documentclass[a4paper]{article}
\usepackage{graphicx}
\usepackage{multirow}

\usepackage{INTERSPEECH2020}
\usepackage[utf8]{inputenc}
\usepackage{tikz,booktabs,float,color,adjustbox}

\title{Relative Positional Encoding for Speech Recognition and Direct Translation}

\name{Ngoc-Quan Pham$^1$~~Thanh-Le Ha$^1$~~Tuan-Nam Nguyen$^1$~~Thai-Son Nguyen$^1$~~Elizabeth Salesky$^2$ Sebastian Stueker$^1$~~Jan Niehues$^3$~~Alex Waibel$^1$$^,$$^4$}
\address{
  $^1$Interactive Systems Lab, Karlsruhe Institute of Technology, Karlsruhe, Germany\\
  $^2$Johns Hopkins University, Baltimore, Maryland \\
  $^3$Department of Data Science and Knowledge Engineering (DKE), Maastricht University\\
  $^4$Carnegie Mellon University, Pittsburgh PA, USA}
\email{ngoc.pham@kit.edu, thanh-le.ha@kit.edu, tuan.nguyen@kit.edu, thai.nguyen@kit.edu\\
esalesky@jhu.edu, sebastian.stueker@kit.edu, jan.niehues@maastrichtuniversity.nl, waibel@cs.cmu.edu}

\begin{document}
\maketitle
\begin{abstract}
Transformer models are powerful sequence-to-sequence architectures that are capable of directly mapping speech inputs to transcriptions or translations.
However, the mechanism for modeling positions in this model was tailored for text modeling, and thus is less ideal for acoustic inputs. 
In this work, we adapt the relative position encoding scheme to the Speech Transformer, where the key addition is relative distance between input states in the self-attention network. 
As a result, the network can better adapt to the variable distributions present in speech data. 
Our experiments show that our resulting model achieves the best recognition result on the Switchboard benchmark in the non-augmentation condition, and the best published result in the MuST-C speech translation benchmark. 
We also show that this model is able to better utilize synthetic data than the Transformer, and adapts better to variable sentence segmentation quality for speech translation. 
\end{abstract}


\section{Introduction}

It is now evident that neural sequence-to-sequence models~\cite{sutskever2014sequence} are capable of directly transcribing or translate speech in an end-to-end approach. 
A single neural model which directly maps speech inputs to text outputs  advantageously eliminates the individual components in non end-to-end or cascaded approaches, while yielding competitive performance~\cite{sperber2019attention,nguyen2019improving}. 
The hybrid approach for speech recognition and the cascaded approach for speech translation may still give the best accuracy in many conditions, but as neural architectures continue to develop, the gap is closing~\cite{niehues_j_2019_3525578}. 

The Transformer~\cite{vaswani2017attention} is a popular architecture choice which has achieved state-of-the-art performance for many sequence learning tasks, particularly machine translation~\cite{ng-etal-2019-facebook}. 
When applied to speech recognition and direct speech translation, this architecture also stands out as the highest performing option for several datasets~\cite{sperber2018self,pham2019transformer,di2019adapting}.

The disadvantage of the Transformer is that, its core function -- \textit{self-attention} -- does not have an inherent mechanism to model \textit{sequential positions}. 
The original work~\cite{vaswani2017attention} added position information to the word embeddings via a trigonometric position encoding. 
Specifically, each element in the sequence is assigned an \textit{absolute} position with a corresponding encoding (a vector similar to embeddings of the discrete variables, but not updated during training). 
Recent adaptation to speech recognition ~\cite{pham2019transformer}\footnote{This is the closest speech adaptation that does not change or introduce additional layers (e.g. LSTM ~\cite{hochreiter1997long} or TDNN~\cite{waibel1989tdnn}).} showed that the base model, extended in depth, is already sufficient for competitive performance compared to other architecture approaches. 

\begin{figure}[htb]
\vspace{-1em}
\centering
\includegraphics[width = 0.45\textwidth]{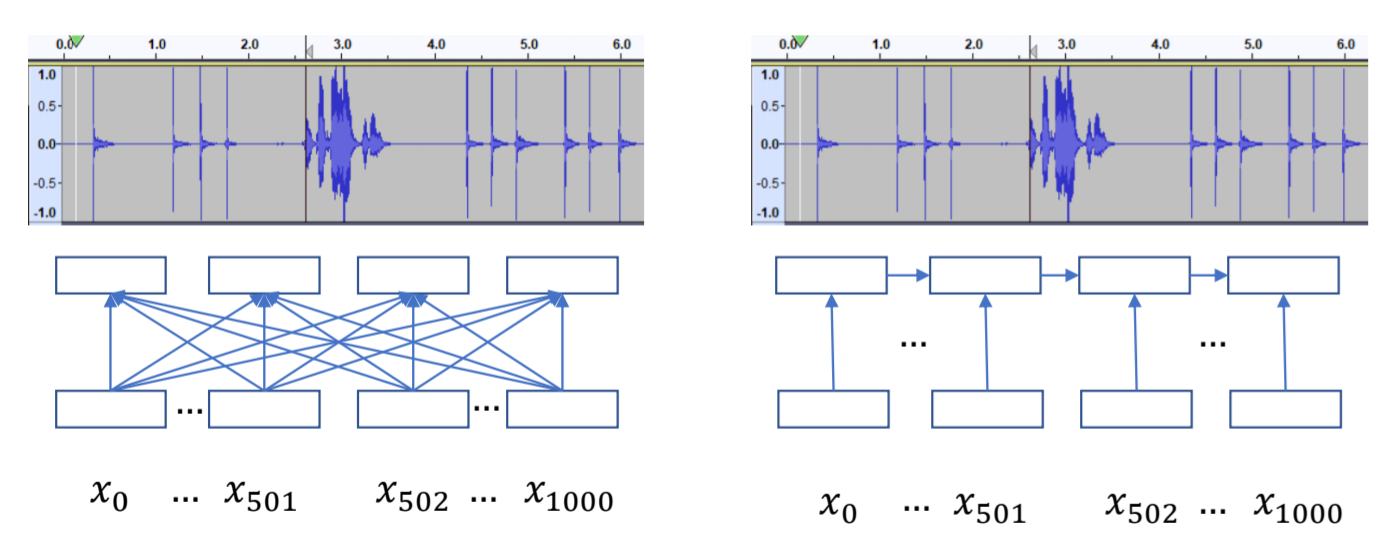}
\caption{\label{fig:problem} Speech utterance with applauses at the start and end. Positions have large variation, which is harmful for Transformer with absolute positions and more computation. LSTMs (right) can alleviate this with the forgetting mechanism.}
\end{figure}

However, this absolute position scheme is far from ideal for acoustic modeling. 
First, text sequences may have a stricter correlation with position; for example, in English the ``Five Ws'' words often appear at the beginning of the sentences, while there may be larger variation in the absolute position of phones in speech signals and utterances. 
Second, speech sequences are often $10-60$ times (in terms of frames) longer than their transcript character sequence, which can be exacerbated by surrounding noises or silences. 
Figure~\ref{fig:problem} shows an illustration in which the speech (at $\approx$ frame 500) is between  applause, which changes absolute positions, but should not affect the resulting transcript.   
Ideally, we want to keep positional information \textit{time-shift invariant}. 

Recently, relative positional encoding has become popularized as a consistent reinforcement for the self-attention. Originally proposed by~\cite{shaw-etal-2018-self} to replace absolute positions by taking into account the \textit{relative positions} between the states in self-attention, this method has also been formalized to adapt into language modeling~\cite{dai-etal-2019-transformer}, which allows the models to capture very long dependency between paragraphs.

In this work, we bring the advantages of relative position encoding to the Deep Transformer~\cite{pham2019transformer} for both speech recognition (ASR) and direct speech translation (ST). 
The resulting novel model maintains the trigonometric position encodings to better scale with longer speech sequences, and is able to model bidirectional positions as well. 
On speech recognition, we show that this model consistently improves the Transformer on the standard English Switchboard and Fisher benchmarks (on both 300h and 2000h conditions), and, to the best of our knowledge, is the best published end-to-end model without augmentation on these datasets. 
More impressively, for speech translation, a single model is able to improve the previous best on the MuST-C benchmark~\cite{mustc19} by $7.2$ BLEU points. 
While extending to the IWSLT speech translation task, which is very challenging because it requires of generating audio segmentations, we find that the relative model scales much better with the segmentation quality than the absolute counterpart, and can challenge a very strong cascaded model, which has the advantage of additional model parameters, an intermediate re-segmentation component, and more data.

\section{Model Description}

A speech-to-text model for either automatic speech recognition or direct translation transforms a source speech input with $N$ frames $X = {x_1, x_2, \dots, x_N}$ into a target text sequence with $M$ tokens $Y = {y_1, y_2, \dots, y_M}$. The encoder transforms the speech inputs into hidden representations $h^X_{1\dots N}$. The decoder firsts generates a language model style hidden representation $h^Y_i$  given the previous inputs, then uses the attention mechanism~\cite{bahdanau2014neural} to generate the relevant context $c_i$ from the encoder states, which is then combined and generate the output distribution $o_i$.

\begin{align}
    h^X_{1\dots N} &= ENCODER(x_1 \dots x_N) \\
    h^Y_i &= DECODER(y_i, y_{1 \dots {i-1}})  \\
    c_i &= \text{ATTENTION}(h^Y_i, h_{1\dots N}) \\
    o_i &= \text{SOFTMAX}(c_i + h^Y_i) \label{eq:combine} \\ 
    y_{i+1} &= sample(o_i) 
\end{align}

\subsection{Transformer}
The Transformer~\cite{vaswani2017attention} uses attention as the main network component to learn encoder and decoder hidden representations. Given three sequences of vectorized states consisting of queries $Q \in R^{\left|{Q}\right| \times D}$, $K, V \in R^{\left|{K}\right| \times D }$, attention computes an energy function $e_{ij}$ between each query $Q_i$ and each key $K_j$. These energy terms are then normalized with a softmax function, and then used to take the weighted average of the values $V$. The energy function can be modeled with neural networks~\cite{luong2015effective} or as simple as projected (with thre additional weight matrices) dot-product between two vectors $e_{ij} = Q_i K^T_j$ or as parallelized matrix-multiplication in Equation~\ref{eq:dot-prod-attn}.

\begin{equation}
\label{eq:dot-prod-attn}
\begin{aligned}
\hat{Q} = QW_Q ; \hat{K} = KW_K ; \hat{V} = VW_V   \\
\text{Attention}(Q, K, V) = \text{softmax}(\hat{Q}\hat{K}^T)V \\
\end{aligned}
\end{equation}

\cite{vaswani2017attention} also improved the attention above through the concept of multi-head attention (MAH), which splits the transformed term $\hat{Q}, \hat{K}, \hat{V}$ to $H$ different heads. The same dot-product operation is applied on each of the $H$ query, key and values heads, and finally the result is the concatenation of the $H$ outcomes.

The Transformer encoder and decoder are constructed based through stacked layers that have identical components. Each encoder layer has one self-attention (MAH) sub-layer, which is followed by a position-wise feed-forward neural network with ReLU activation function.\footnote{It is a sub-layer from the top-down perspective, analyzing the network, but as a neural network itself, it has two hidden layers of its own} 
Each decoder layer is quite similar to the encoder counterpart, with the self-attention sub-layer to connect the decoder states, and the feed-forward network. There is an addition encoder-decoder attention layer in between to extract the context vectors from the top encoder states. 
Furthermore, the Transformer uses residual connections boost information from bottom layers (e.g. the input embeddings) to the top layers. 
Layer normalization~\cite{ba2016layer} plays a supportive role, keeping the norms of the outputs in check, when used after each residual connection.  

\subsection{Relative Position Encoding in Transformer}
Equation~\ref{eq:dot-prod-attn} suggests that attention is position-invariant, i.e if the key and value states change their order, the output remains the same. 
In order to alleviate this problem for this content-based model, positional information within the input sequence is represented in a similar manner with the word embeddings. The positions are treated as discrete variables and then transformed to embeddings either using a look-up table with learnable parameters~\cite{sukhbaatar2015end} or with fixed encodings in a trigonometric form: 

\begin{equation}
\begin{aligned}
    P{i, 2k} &= sin(\frac{i}{10000^{2k/D}}) \\
    P{i, 2k+1} &= cos(\frac{i}{10000^{2k/D}})
\end{aligned}
\end{equation}

When applied to \textit{speech} input, this encoding is then added to speech input features~\cite{pham2019transformer}. The periodic property of the encodings allow the model to generalize to unseen input length. Following the factorization in~\cite{dai-etal-2019-transformer}, we can rewrite the energy function in Equation~\ref{eq:dot-prod-attn} for self-attention between two encoder hidden states $H_i$ and $H_j$ to decompose into 4 different terms:

\begin{equation}
\label{eq:factorize}
\begin{aligned}
\text{Energy}_{ij} &=  Energy(H_i + P_i, H_j + P_j) \\
     &=  H_iW_QW_K^TH_j^T + H_iW_QW_K^TP_j^T \\ 
     & \quad  + P_iW_QW_K^TH_j^T + P_iW_QW_K^TP_j^T \\
     &= A + B + C + D
\end{aligned}
\end{equation}

 Equation~\ref{eq:factorize} gives us an interpretation of the function: in which term A is purely content-based comparison between two hidden states (i.e speech feature comparison), term D gives a bias between two absolute positions. The other terms represent the specific content and position addressing. 

The extension proposed by previously~\cite{shaw-etal-2018-self} and later~\cite{dai-etal-2019-transformer} changed the terms B, C, D so that only the relative positions are taken into account:

\begin{equation}
\label{eq:relative}
\begin{aligned}
\text{Energy}_{ij} &=  Energy(H_i, H_j + P_{i-j}) \\
     &=  H_iW_QW_K^TH_j^T + H_iW_QW_R^TP_{i-j}^T \\ 
     & \quad +~uW_K^TH_j^T +~ vW_R^TP_{i-j}^T \\
     &= A + \tilde{B} + \tilde{C} + \tilde{D}
\end{aligned}
\end{equation}

The new term $\tilde{B}$ computes the relevance between the input query and the relative distance between $Q$ and $K$. Term  $\tilde{C}$ introduces an additional bias $v$ to the content of the key state $H_j$, while term $\tilde{D}$ represents the bias to the global distance. Terms $\tilde{B}$ and $\tilde{D}$ also have an additional linear projection $W_R$ so that the positions and embeddings have different projections. 

With this relative position scheme, when the two inputs $H_i$ and $H_j$ are shifted (for example, having extra noise or silent in the utterance), the energy function stays the same (for the first layer of the network). 
Moreover, it can also establish certain inductive bias in the data; for example, the average length of silence or applauses, given the global and local bias terms. 

\subsection{Adaptation to speech inputs}

For relative position encodings with speech inputs, should we use learnable embeddings or fixed encodings to represent the distance $P_i$? 
The latter has the clear advantage that it already has the periodic property, and given that speech input can be as long as thousands of frames, the former approach would require a necessary cut-off~\cite{shaw-etal-2018-self} to adapt to longer input sequences. These reasons make sinusoidal encodings a logical choice.  

Importantly, the relative position scheme above was proposed for autoregressive language models, in which the attention has only one direction. For speech encoders, each state can attend to both left and right directions, thus we propose to use positive distance when the keys are to the left ($j < i$) and negative distance otherwise. As a result, the encodings for $P_k$ and $P_{-k}$ will have the same $sin$ terms while the $cos$ terms will have opposite signs, which gives the model a hint to assign different biases to different directions. Implementation wise, it is able to efficiently compute terms $\tilde{B}$ and $\tilde{D}$ with the minimal amount of matrix operations. It is necessary to compute $2K-1$ terms $H_iW_QW_R^TP_{k}^T$ with $-K < k < K$ for each query $H_i$ (For a sequence with $K$ states, the distance between one state to another $k$ is always in that range).\footnote{\cite{dai-etal-2019-transformer} only needs to compute $K$ terms as it has only one direction} 
This is followed by the shifting trick~\cite{dai-etal-2019-transformer} to achieve the required energy terms.

\section{Experiments}

\subsection{Datasets}

\underline{\textbf{ASR}}. For ASR tasks, our experiments were conducted on the standard English Switchboard and Fisher data under both benchmark conditions: $300$ hours and $2000$ hours of training data. Our reported test results are for the Hub5 testset with two subsets Switchboard and CallHome. Target transcripts are segmented with byte-pair encoding \cite{sennrich2016bpeacl} using $10$k merges. 

~\\
\noindent\underline{\textbf{SLT}}. We split our SLT task into two different subtasks. Many SLT datasets require an auto-segmentation component to splits the audio into sentence-like segments.\footnote{This is commonly seen in IWSLT evaluation campaigns~\cite{niehues_j_2019_3525578}.} 
For end-to-end models, this step is crucial due to the lack of incremental decoding and higher GPU memory requirements. The recent~\emph{MuST-C}~\cite{mustc19} corpus contains segmentations for both training and testset, requiring no extra segmentation component, and so we use its English-German pair serves as our first experimental benchmark. 
We further carry out experiments on the IWSLT 2019 evaluation campaign data, a superset of MuST-C, where segmentation is not given; here we can compare the effects of variable-quality segmentation on different end2end models, and also compare models to highly competitive tuned cascades. 
We use the MuST-C validation data for both tasks.

\subsection{Setup}
Our baselines for all experiments use the Deep Stochastic Transformer~\cite{pham2019transformer}. We use the relative encoding scheme above for both encoder and decoder to yield relative Transformers.

For ASR, both our baseline Transformer and relative Transformer have $36$ encoder and $12$ decoder layers with the model size $D=512$ and the feed-forward networks have the hidden layer size of $2048$. Dropout is applied with the same mask across time steps~\cite{gal2016theoretically} with $P_{drop}=0.35$ and also directly at the discrete decoder inputs with $P_{drop}=0.1$.  All models are trained for at most $120000$ steps and the reported model parameters are the average of the 10 checkpoints with lowest perplexities on the cross-validation data.

For SLT, the models and the training process are identical to ASR, with the exception that we use $32$ encoder layers.\footnote{The SLT data sequences are longer and thus need more memory} 
Following the curriculum learning intuition that SLT models benefit from pre-training the speech encoder with ASR~\cite{bansal2018pre}, we first pre-trained the model for ASR with the parallel English transcripts from MuST-C, and then fine-tune the encoder weights and re-initialize the decoder for SLT. This approach enabled us to consistently train our SLT models without divergence (which may happen when the learning rate is too aggressive or the half-precision GPU mode is used). 

For all models, the batch size is set to fit the models to a single GPU~\footnote{Titan V and Titan RTX with 12 and 24 GB respectively} and accumulate gradients to update every $12000$ target tokens. We used the same learning rate schedule as the Transformer translation model~\cite{vaswani2017attention} with $4096$ warmup steps for the Adam~\cite{kingma2014adam} optimizer. 

\subsection{Speech Recognition Results} 

\vspace{-1em}
\begin{table}[ht]
\caption{ASR: Comparing our best models to other hybrid and end-to-end systems on the \textbf{300h} SWB training set and Hub5'00 test sets. Absolute best is bolded, our best is italicized. WER$\downarrow$~. }
\label{tab:swb1}
	\vspace{-0.2cm}	
	\setlength{\tabcolsep}{4pt}
	\centering
	\begin{adjustbox}{width=\columnwidth}
	\begin{tabular}{llcccc}
		\toprule
        & \textbf{Models} & \textbf{SWB} & \textbf{w/ SA} & \textbf{CH} & \textbf{w/ SA} \\
        \midrule
        \parbox[t]{2mm}{\multirow{2}{*}{\rotatebox[origin=c]{90}{\textbf{Hyb.}}}} & \cite{povey2016purely} BLSTM+LFMMI  & 9.6  & -- & 19.3 & -- \\
        & \cite{zeyer2018improved} Hybrid+LSTMLM & 8.3 & -- & 17.3 & -- \\
        \midrule
        
        \parbox[t]{2mm}{\multirow{5}{*}{\rotatebox[origin=c]{90}{\textbf{End-to-End}}}} &  \cite{park2019specaugment} LAS (LSTM-based) &  11.2 & \textbf{7.3} & 21.6 & \textbf{14.4}    \\
        & \cite{zeyer2019comparison} Shallow Transformer & 16.0 & 11.2 & 30.5 & 22.7 \\
        & \cite{zeyer2019comparison} LSTM-based & 11.9 & 9.9 & 23.7 & 21.5 \\
        & \cite{nguyen2019improving} ~~LSTM-based & 12.1 & 9.5 & 22.7 & 18.6 \\
        & \quad +SpecAugment +Stretching & -- &\textit{8.8} & -- & 17.2 \\
        \midrule
        \parbox[t]{2mm}{\multirow{4}{*}{\rotatebox[origin=c]{90}{\textbf{Ours}}}} & Deep Transformer \textit{(Ours) }  & 10.9 & 9.4 & 19.9 & 18.0 \\
        & \quad +SpeedPerturb & -- & \textit{9.1} & -- & \textit{17.1} \\ 
        & Deep Relative Transformer \textit{(Ours)} & \textbf{10.2} & 8.9 & 19.1 & 17.3 \\
        & \quad +SpeedPerturb  & -- & \textit{8.8} & -- & \textbf{16.4} \\
		\bottomrule
	\end{tabular}
	\end{adjustbox}
\end{table}

We present ASR results on the Switchboard-300 benchmark in Table~\ref{tab:swb1}. It is important to clarify that spectral augmentation (dubbed as \textit{SpecAugment}) is a recently proposed augmentation method that tremendously improved the regularization ability of seq2seq models for speech recognition~\cite{park2019specaugment}. In better demonstrate the effect of relative attention, we conduct experiments with and without augmentation.

Compared to the Deep Stochastic model~\cite{pham2019transformer}, using relative attention is able to reduce our WER from $10.9$ to $10.2$ and $19.9$ to $19.1$ on SWB and CH, without any augmentation. 
Compared to other works under this condition, our results are second to none among the published end2end models, and can rival the LFMMI hybrid model~\cite{povey2016purely} that has an external language model  utilizing extra monolingual data. 

With spectral augmentation, the improvement from relative attention is still noticeable, further reducing WER from $9.4$ to $8.9$, and $18.0$ to $17.3$ from the baseline (the relative gain on \textbf{CallHome} is kept at $4\%$). This is second only to~\cite{park2019specaugment}, the state-of-the-art on this benchmark at $7.3$ and $14.4$; however their models use an aggressively regularized training regime on multiple TPUs for 20 days. Other end-to-end models~\cite{zeyer2019comparison,nguyen2019improving} using single GPUs showed similar behavior to ours with SpecAugment. Finally, with additional speed augmentation, relative attention is still additive, with further gains of 0.3 and 0.7 compared to our strong baseline. 

\begin{table}[ht]
\caption{ASR: Comparison on \textbf{2000h} SWB+Fisher training set and Hub5'00 test sets. Absolute best is bolded, our best is italicized. WER$\downarrow$~.}
\label{tab:swb2}
	\vspace{-0.2cm}	
	\setlength{\tabcolsep}{4pt}
	\centering
	\begin{tabular}{llcc}
		\toprule
        & \textbf{Models} & \textbf{SWB} & \textbf{CH}\\
        \midrule
        \parbox[t]{2mm}{\multirow{3}{*}{\rotatebox[origin=c]{90}{\textbf{Hybrid}}}} & \cite{povey2016purely} Hybrid  & 8.5 & 15.3 \\
        & \cite{saon2017english} Hybrid w/ BiLSTM  & 7.7 & 13.9 \\
        & \cite{han2017capio} Dense TDNN-LSTM  & \textbf{6.1} & \textbf{11.0} \\
        \midrule
        \parbox[t]{2mm}{\multirow{4}{*}{\rotatebox[origin=c]{90}{\textbf{End-to-End}}}} &  \cite{audhkhasi2018building} CTC & 8.8 & 13.9 \\
        & \cite{nguyen2019improving}~~ LSTM-based & 7.2 & 13.9 \\
        		\vspace{-0.2cm}  \\
        & Deep Transformer \textit{(Ours)}   & 6.5 & 11.9 \\
        & Deep Relative Transformer \textit{(Ours)} & \textit{6.2} & \textit{11.4} \\
		\bottomrule
	\end{tabular}
\end{table}

The experiments on the larger dataset with 2000h follow the above results for 300h, continuing to show positive effects from that relative position encodings. 
The error rates on those SWB and CH decrease from $6.5$ and $11.9$ to $6.2$ and $11.4$ (Table \ref{tab:swb2}). 
Our best model is significantly better than previously published CTC~\cite{audhkhasi2018building} and LSTM-based~\cite{nguyen2019improving} models, and approaches the heavily tuned hybrid system~\cite{han2017capio} with dense TDNN-LSTM. 
It is likely possible to reach better error rates, with the help of ensembled models, further data augmentation, and language models. 
Our experiments here, however, show that the novel relative model is consistently better than the baseline, regardless of the data size and augmentation conditions. 

\subsection{Speech Translation Results}

Our first SLT models were trained only on the MuST-C training data and the results are reported on the COMMON testset\footnote{MuST-C is a multilingual dataset and this testset is the commonly shared utterances between the languages.}. Note that in this testset, we are provided with the segmentation of each utterance which has a corresponding translation. For each utterance, we can directly translate with the end2end model, and the final score can be obtained using standard BLEU scorers such as SacreBLEU~\cite{post-2018-call} because the output and the reference are already sentence-aligned in a standardized way. 

As shown in Table~\ref{tab:slt1}, our Deep Transformer baseline achieves an impressive $24.2$ BLEU score compared to the ST-Transformer~\cite{di2019adapting}, which is a Transformer model specifically adapted for speech translation. However, using relative position information makes self-attention more robust and effective still, as our BLEU score increases to $25.2$. 

To try to maximize the performance of an end-to-end speech translation model, we also add the  Speech-Translation TED corpus~\footnote{Available from the evaluation campaign at https://sites.google.com/view/iwslt-evaluation-2019/speech-translation} and follow the method from~\cite{di2019adapting} to add synthetic data for speech translation, where a cascaded system is used to generate translations for the TEDLIUM-3 data~\cite{hernandez2018ted}. Our cascade system is built based on the procedure from the winning system in the 2019 IWSLT ST evaluation campaign~\cite{pham2019iwslt}.

With these additional corpora, we observe a considerable boost in translation performance (similarly observed in~\cite{di2019adapting}). More importantly, the relative model further enlarges the performance gap between two models to now $1.4$ BLEU points. We hypothesize that the model is able to more effectively use the additional data, with data patterns more easily captured when the model considers \textit{relative} rather than absolute distance between speech features. More concretely, each training corpus has a different segmentation method, which leads to large variation in spoken patterns, which is difficult to capture using absolute position encodings.

To verify our hypothesis, we compare these two models and the cascaded system on the TEDTalk testsets without a provided segmentation. These talks are available as long audio files and require an external audio segmentation step to make translation feasible. It is important to note that the cascaded model has a separate text re-segmentation component~\cite{cho2017nmt} which takes ASR output and reorganizes it into logical sentences, which is a considerable advantage compared to the end2end models. We experimented with several audio segmentation methods and see that the cascade is less affected by the segmentation quality than the end-to-end models.

The results in Table~\ref{tab:slt2} compare two different segmentation methods, LIUM~\cite{rouvier2013open} and VAD~\cite{Charles2013}, and four different testsets. 
The relative Transformer unsurprisingly consistently outperforms the Transformer, regardless of segmentation. 
Moreover, comparing between the segmenters, the relative model more effectively uses higher segmentation quality, yielding a larger BLEU difference. 
While the base Transformer only increases up to $0.5$ BLEU with better segmentation, this figure becomes up to $2.4$ BLEU points for the relative counterpart. In the end, the cascade model still shows that heavily tuned separated components, together with an explicit text segmentation module, is an advantage over end-to-end models, but this gap is closing with more efficient architectures. 
\vspace{-0.5em}
\begin{table}[ht]
\caption{ST: Translation performance in BLEU$\uparrow$ on the COMMON testset (no segmentation required) }
\label{tab:slt1}
	\vspace{-0.2cm}	
	\setlength{\tabcolsep}{4pt}
	\centering
	\begin{tabular}{lc}
		\toprule
        \textbf{Models} & BLEU \\
        \midrule
        \cite{di2019adapting} ST-Transformer & 18.0 \\
        \quad +SpecAugment  & 19.3 \\
        \quad +Additional Data \cite{di2019data} & 23.0 \\
        \midrule
        Deep Transformer (w/ SpecAugment) & 24.2 \\
        \quad +Additional Data & 29.4 \\
        Deep Relative Transformer (w/ SpecAugment) & \textbf{25.2} \\
        \quad +Additional Data & \textbf{30.6} \\
		\bottomrule
	\end{tabular}
	\vspace{-0.0cm}
\end{table}

\vspace{-1em}
 \begin{table}[ht!]
 \caption{ST: Translation performance in BLEU$\uparrow$ on IWSLT testsets (re-segmentation required) }
   \label{tab:slt2}
   	\vspace{-0.2cm}	
  \centering
  	\begin{adjustbox}{width=\columnwidth}
  \begin{tabular}{lcccccc}
  \toprule
  \textit{Testset} \quad ~~$\rightarrow$ & \multicolumn{2}{c}{\textbf{Transformer}} & \multicolumn{2}{c}{\textbf{Relative}} & \multicolumn{2}{c}{\textbf{Cascade}} \\
  \textit{Segmenter} $\rightarrow$ & LIUM & VAD & LIUM & VAD & LIUM & VAD  \\
  \cmidrule(lr){1-1} \cmidrule(lr){2-3} \cmidrule(lr){4-5} \cmidrule(lr){6-7}
  tst2010 & 22.04   & 22.53    &  23.29  & 24.27   & 25.92  & 26.68 \\
  tst2013 & 25.74   & 26.00    &  27.33  & 28.13   & 27.67  & 28.60 \\
  tst2014 & 22.23   & 22.39    &  23.00  & 25.46   & 24.53  & 25.64 \\
  tst2015 & 20.20   & 20.77    &  21.00  & 21.82   & 23.55  & 24.95 \\
  \bottomrule
  \end{tabular}
  \end{adjustbox}
  \vspace{-1em}
\end{table} 

\section{Conclusion}
Speech recognition and translation with end-to-end models have become active research areas. In this work, we adapted the relative position encoding scheme to speech Transformers for these two tasks. We showed that the resulting novel network provides consistent and significant improvement through different tasks and data conditions, given the properties of acoustic modeling. Inevitably, audio segmentation remains a barrier to end-to-end speech translation; we look forward to future neural solutions. 


\bibliographystyle{IEEEtran}

\bibliography{mybib}


\end{document}